\documentclass[conference]{IEEEtran}
\IEEEoverridecommandlockouts
\usepackage{cite}
\usepackage{amsmath,amssymb,amsfonts}
\usepackage{algorithmic}
\usepackage{graphicx}
\usepackage{wrapfig}
\usepackage{amssymb}

\usepackage{subcaption}

\usepackage{textcomp}
\usepackage{amsmath}
\usepackage{xcolor}
\def\BibTeX{{\rm B\kern-.05em{\sc i\kern-.025em b}\kern-.08em
    T\kern-.1667em\lower.7ex\hbox{E}\kern-.125emX}}
\begin{document}
\bstctlcite{IEEEexample:BSTcontrol}

\title{Scalable Coherent Optical Crossbar Architecture using PCM for AI Acceleration \\
\thanks{Acknowledgement Place Holder}
}

\author{\IEEEauthorblockN{Dan Sturm}
\IEEEauthorblockA{
\textit{Electrical and Computer Engineering} \\
\textit{University of Washington}\\
Seattle, USA \\
dansturm@uw.edu}
\and
\IEEEauthorblockN{Sajjad Moazeni}
\IEEEauthorblockA{
\textit{Electrical and Computer Engineering} \\
\textit{University of Washington}\\
Seattle, USA \\
smoazeni@uw.edu}
}


\maketitle

\begin{abstract}

Optical computing has been recently proposed as a new compute paradigm to meet the demands of future AI/ML workloads in datacenters and supercomputers. However, proposed implementations so far suffer from lack of scalability, large footprints and high power consumption, and incomplete system-level architectures to become integrated within existing datacenter architecture for real-world applications. In this work, we present a truly scalable optical AI accelerator based on a crossbar architecture. We have considered all major roadblocks and address them in this design. Weights will be stored on chip using phase change material (PCM) that can be monolithically integrated in silicon photonic processes. All electro-optical components and circuit blocks are modeled based on measured performance metrics in a 45nm monolithic silicon photonic process, which can be co-packaged with advanced CPU/GPUs and HBM memories. We also present a system-level modeling and analysis of our chip's performance for the Resnet-50V1.5, considering all critical parameters, including memory size, array size, photonic losses, and energy consumption of peripheral electronics. Both on-chip SRAM and off-chip DRAM energy overheads have been considered in this modeling. We additionally address how using a dual-core crossbar design can eliminate programming time overhead at practical SRAM block sizes and batch sizes. Our results show that a $128\times128$ proposed architecture can achieve inference per second (IPS) similar to Nvidia A100 GPU at $15.4 \times$ lower power and $7.24 \times$ lower area.

\end{abstract}


\section{Introduction}
\label{sec:intro}

\begin{figure}[t]
    \begin{center}
        \includegraphics[width=0.9\linewidth]{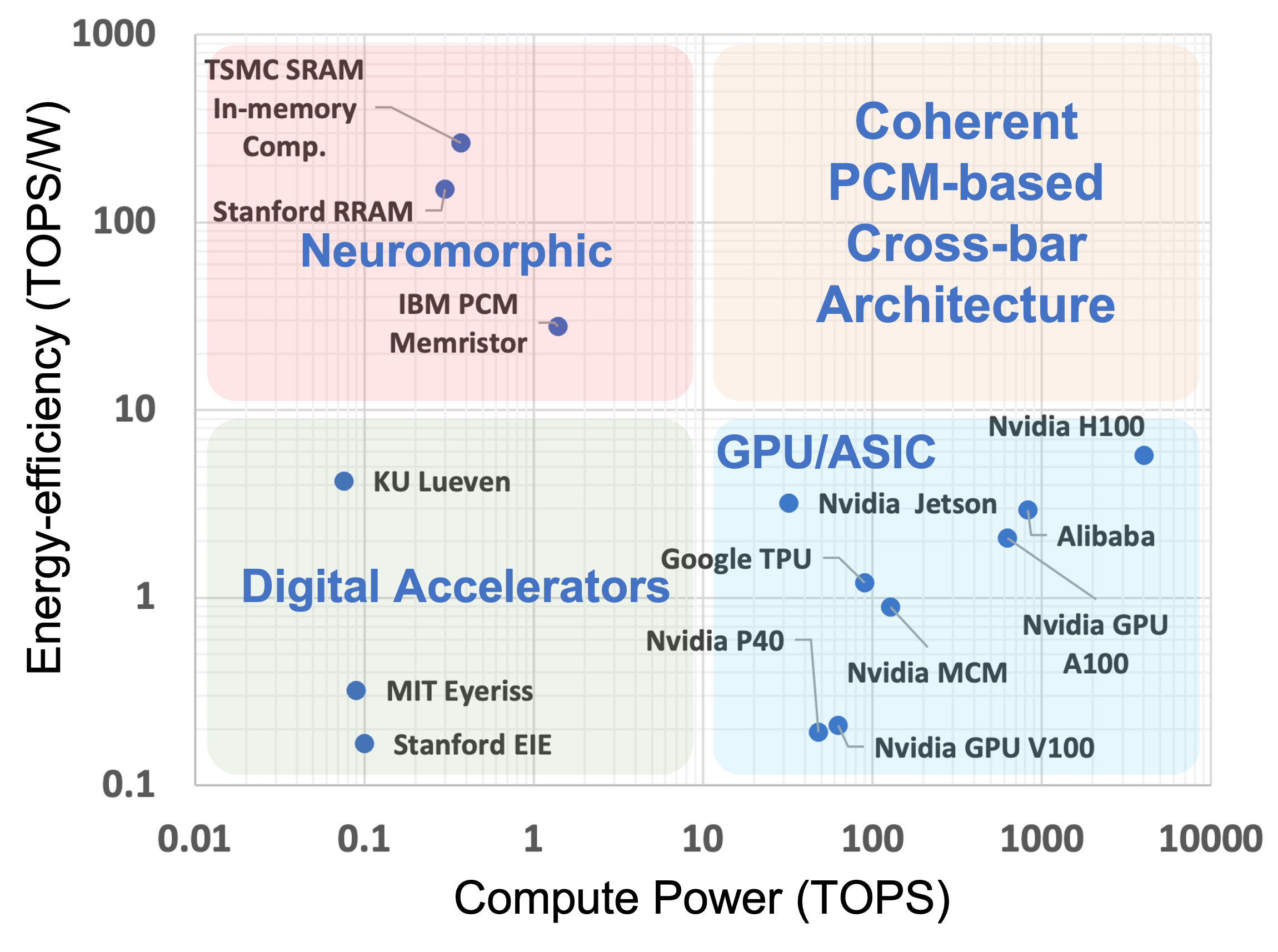}
        \vspace*{-0.5\baselineskip}
        \caption{Comparison of state-of-the-art AI/ML processors.}
        \label{fig:TOPS}
    \end{center}
    \vspace*{-2\baselineskip}
\end{figure}

Recent advancements in artificial intelligence (AI) and machine learning (ML) have been challenging our conventional computing paradigms by demanding enormous computing power at a dramatically faster pace than Moore’s law~\cite{Amodei2018}. We can compare the performance of today's AI/ML processors from two key aspects of compute power in terms of Tera operations per second (TOPS) and energy-efficiency (TOPS/W) as illustrated in Fig.~\ref{fig:TOPS}. Despite the promising success of neuromorphic and analog-based computing in electrical domains for low TOPS applications (edge-computing), these approaches can not satisfy the requirements of datacenters and super computers. Due fundamental bandwidth limitations, they cannot achieve high throughput. Optical neural networks (ONNs) can potentially overcome this barrier by providing tens of GHz bandwidths and ultra-low losses of photonic integrated circuits~\cite{Shen2017,Xu2021}. However, realizing a practical ONN-based AI accelerator requires a holistic system design that considers devices, circuits, chip architectures, and algorithms.

In this paper, we present a novel architecture for an ONN accelerator in which the multiply-and-accumulate (MAC) operation is performed on a coherent photonic crossbar with programmable phase-change materials (PCM). PCM enables low power photonic computing by storing the weights on-chip in a nonvolatile fashion. This design provides a compact and scalable solution for the first time, which minimally relies on thermo-optic phase-shifters effect. We have considered all critical circuit blocks and parameters, including memory size, array size, photonic losses, and energy consumption of peripheral electronics including analog-to-digital converters (ADCs), digital-to-analog converters (DACs), serializers, and clocking. We develop a custom simulation framework based on existing cycle-accurate simulation tools to model the compute cycles, programming cycles, and DRAM accesses for a given neural network running on a specific set of parameters (including size of SRAM, array, and batch).

Presented work focuses on inference of convolutional neural networks (CNN) such as ResNet50 v1.5, which is used as a benchmark to compare our proposed accelerator performance with state-of-the-art. Additionally, while precision and process variation are major factors in all analog-based computers, here we assume a INT6 precision for all the components as it has been shown to be sufficient for neural networks with high accuracy~\cite{NEURIPS2018_8bit,NEURIPS2020_8bit}. 

This paper is organized as follows: We briefly describe related work in Section~\ref{sec:related-work}. In Section~\ref{sec:xbar-mvm}, we explain the principles of performing the MAC operation in this work. We present an overview of overall chip architecture and CNN operation in Section~\ref{sec:chip-arch}, and section~\ref{sec:sim-framework} explains our custom simulation methodology. Finally, we present the results of our fully optimized design in Section~\ref{sec:results}, and compare those with state-of-the-art in Section~\ref{sec:conclusion}.

\section{Related Work}
\label{sec:related-work}

Researchers have recently proposed a variety of methods to realize an ONN. Most of these works focus only on the physics and devices rather than providing a system-level solution and analysis. In this presented discussion, we only consider integrated solutions, as free-space solutions~\cite{ozcan-science2018} lack the reconfigurability that is an essential part of any “computer” and compatibility with mainstream CMOS technology. Furthermore, we note that a suitable application space for ONNs can be datacenters as opposed to edge computing according to Fig.~\ref{fig:TOPS}. Below we discuss the most promising solutions so far from the perspective of three critical factors that we believe have been addressed in this work:  

\textbf{(1) Scalability:} While on-chip photonics provide high bandwidths, their footprints are fundamentally orders of magnitude larger than advanced nm-scale CMOS. With only one or two routing layers, building an ONN processor with big dimensions has remained elusive. This will become even more challenging considering the need for compact ADCs and DACs. Mach-Zehnder Interferometer (MZI)-based coherent architectures such as~\cite{Shen2017} have large chip areas and large-scale realizations end up exceeding a few $cm^2$. Non-coherent PCM-based crossbars have been also proposed~\cite{Feldmann2021}, however they require many wavelengths for large matrix operations and that is impractical. Time-multiplexed coherent arrays~\cite{nathan-crossbar,Hamerly2019-LargeScaleON} also require 2D arrays of free-space detectors and yet only perform vector-vector multiplication in one clock cycle.

\textbf{(2) Monolithic Integration:} Since any practical computing system will eventually require high-density CMOS electronics for I/O and memory, we argue that electronics and photonic should be monolithically integrated on a single chip using processes such as GF 45CLO~\cite{45CLO-OFC}. While 3D integration is typically proposed as an alternative, existing advanced 3D integration technologies (micro-bumps at $55\mu m$ pitch~\cite{}) cannot provide density for optical computing applications.

\textbf{(3) Full Architecture-level Modeling and Optimization:} Practical AI accelerators, including optical processors, should be modeled at the system level. This has been previously discussed in~\cite{lightmatter-arch}, however, accessing DRAM through a PCIe switch will have large energy and latency overheads. We elaborate on this aspect here, and model the system with co-packaged high-bandwith memory (HBM), similar to state-of-the-art AI accelerators. In addition, we discuss impacts and trade-offs between the programming time, batch size, and multiple cores in this work.

\section{Proposed Crossbar Design for Optical MAC}
\label{sec:xbar-mvm}
The overall proposed crossbar ONN is shown in Fig.~\ref{fig:extended_crossbar}. Below, we describe two key components of this design:

\begin{figure}[t]
    \begin{center}
        \includegraphics[width=0.9\columnwidth]{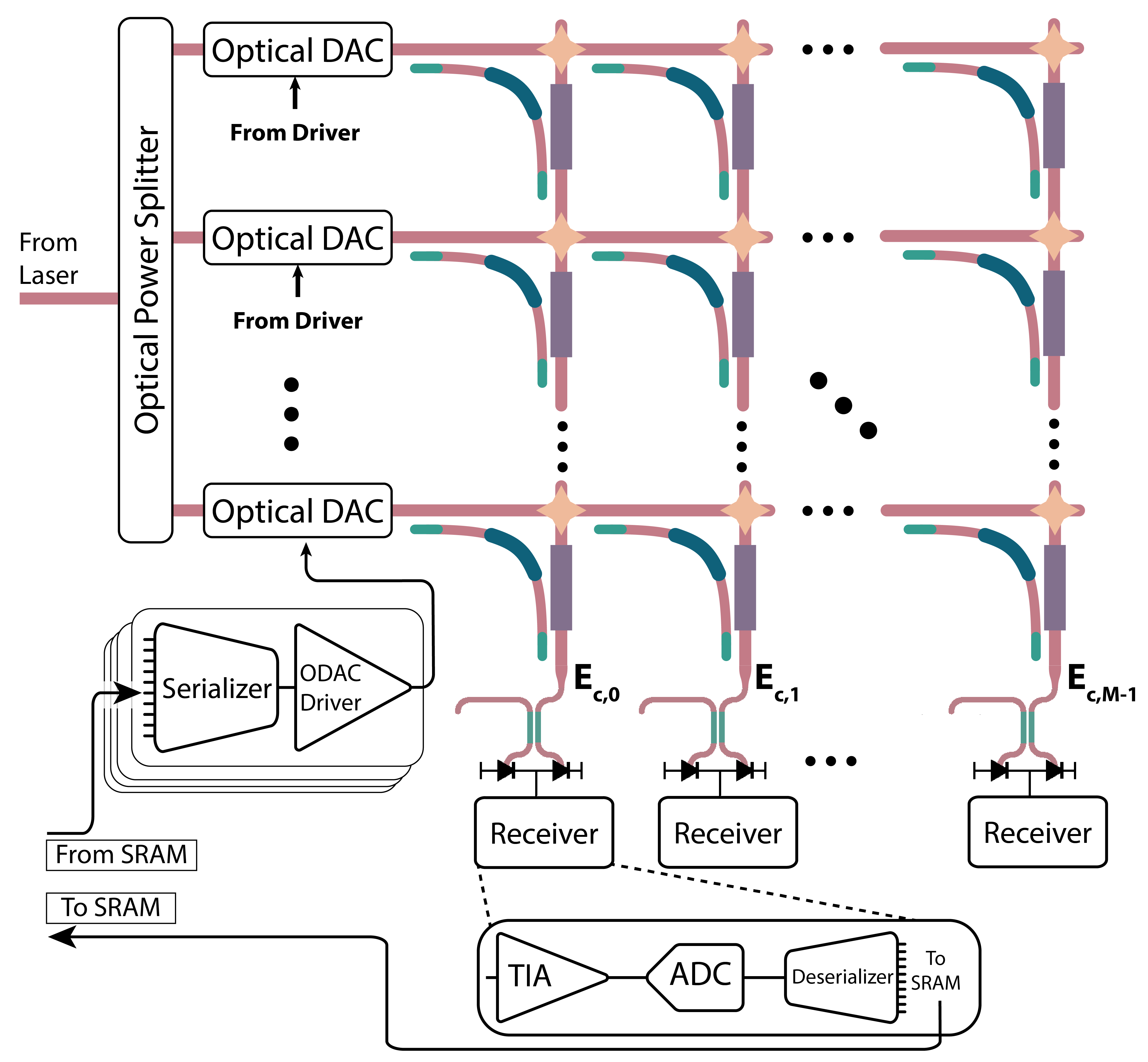}
        \caption{Photonic crossbar array with peripheral electronics for transmitter and receiver}
        \label{fig:extended_crossbar}
    \end{center}
    \vspace*{-2\baselineskip}
\end{figure}

\subsection{Analog-based Optical MAC Core}
The cross-bar design is an $N \times M$ array of PCM-based unit cells. Details of these unit cells and how the array perform MAC is briefly described below.

\subsubsection{PCM-Based Unit Cell}
\begin{figure}[t]
    \begin{center}
        \includegraphics[width=0.55\columnwidth]{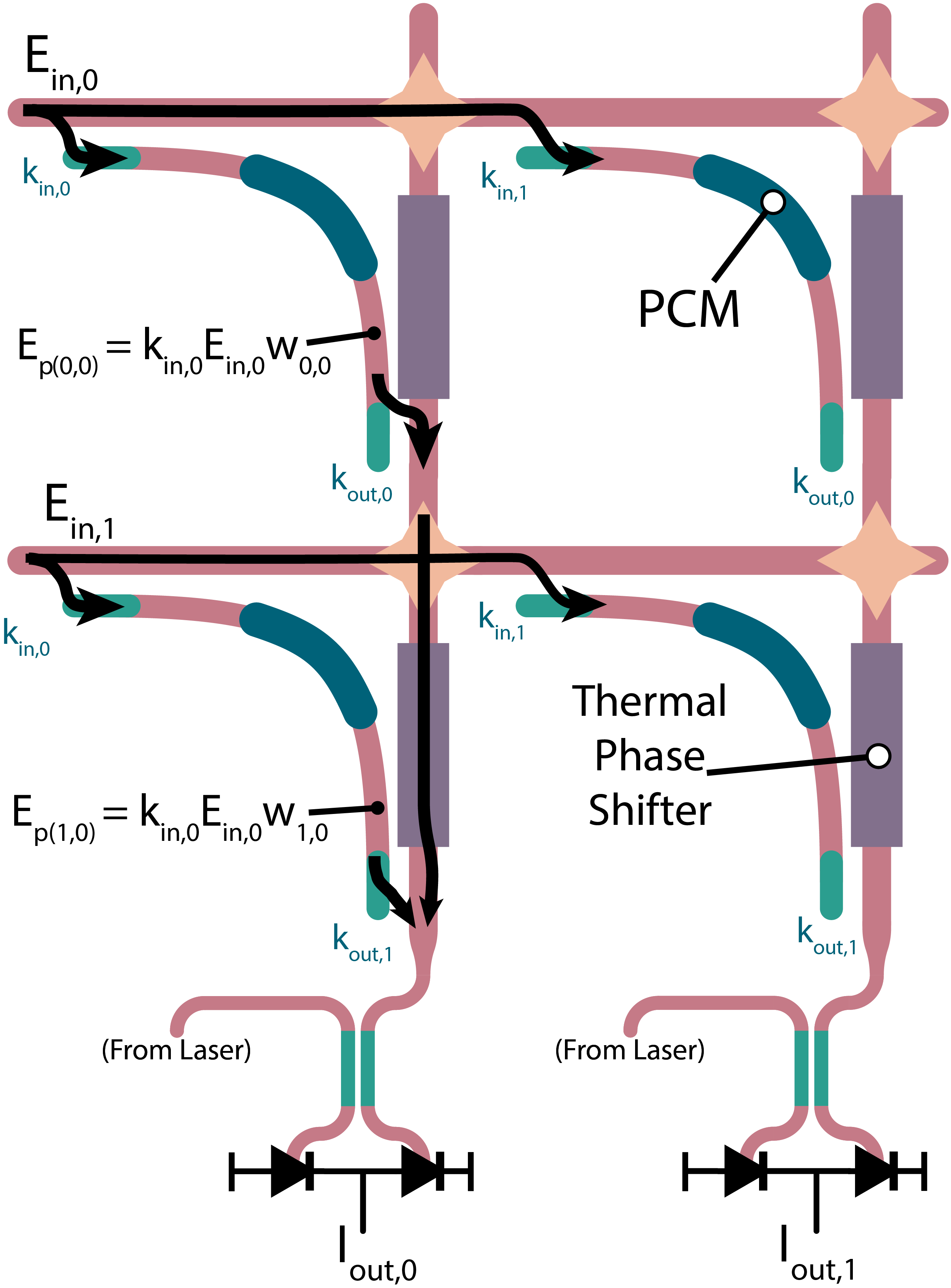}
        \caption{Principle of optical MAC operation in a coherent crossbar array (an example of $2\times 2$ unit cell array).}
        \label{fig:basic_xbar}
    \end{center}
    \vspace*{-2\baselineskip}
\end{figure}

Each unit cell multiplies an input electric field (E-field) by the weight programmed into the PCM section, and adds this product to an externally-inputted electric field through each column. To do so, a portion of light from the row waveguide (E-field into each row is denoted by $|E_{in,i}|$ in Fig.~\ref{fig:basic_xbar}) is partially coupled into a bended waveguide via a directional coupler (DC) with a cross-coupling ratio of $k_{in,j}$ (input coupling strength is column-dependent). The portion of the E-field that does not couple into the unit cell passes through a multi-mode interference (MMI) waveguide crossing junction and enters the next column of unit cells to the right. Each bended waveguide has a $\mu m$-long section covered with PCM. Individual PCM cells can be programmed electrically to be either in the amorphous or crystalline state, or somewhere in between, in a non-volatile fashion~\cite{Feldmann2021,nathan-crossbar}. Programming energy is estimated to be around $100pJ$~\cite{Feldmann2021,nathan-crossbar}. The state of PCM changes the absorption coefficient, and hence it can change the amplitude of E-field. Consequently, if the PCM's programmed transmission in E-field domain is $w_{i, j}$, the E-field at the end of each bended waveguide will be $|E_{in, i}| \times  k_{in,j} \times w_{i, j}$, which we refer to as $E_{p,(i, j)}$. This outcome will be coupled into a column waveguide via another DC (with $k_{out,i}$ coupling ratio), and it will travel into the next row in the bottom, where the coherent summation with another row of products will occur through the DC region across each column. The output coupling strength is row-dependent.

\subsubsection{Crossbar Array of Unit Cells}
In order to perform a MAC operation on our $N \times M$ crossbar array (Fig.~\ref{fig:basic_xbar}), we need to encode the input data vector ($v_{in}$) on each row's E-field amplitude. To do so, we use a splitter tree structure and a set of optical DACs (ODACs) to generate $v_{in,i} \times E_{Laser}/\sqrt{N}$ at each row. Correctly designed input coupling coefficients in each unit cell ($k_{in,j}$) distribute this equally throughout the row, so each unit cell receives $v_{in,i} \times E_{Laser}/\sqrt{NM}$ in its bended waveguide. 

Similarly, output coupling coefficients ($k_{out,i}$) are designed correctly so that outputted light equally represents the product computed by every unit cell in a given column. Although this adds an electrical field loss of $1/\sqrt{N}$, this enables the entire array to operate on a single frequency of light in a compact footprint (unlike previous works like~\cite{Shen2017}), which is critical for scaling the array for high-performance compute. We can calculate the resultant E-field at the end of each column as:

\begin{equation}
 E_{j}^c = \frac{E_{Laser}}{N\sqrt{M}} \sum_{i=0}^{N-1} |v_{in,i}| \times w_{i,j}
\end{equation}



Thus, the overall $E^c$ vector will be equal to the matrix-vector multiplication of input data ($v_{in}$) by a weights matrix of $w$. We can convert this back into the electrical domain using coherent detection by coupling each signal with a certain portion of the input laser light in a DC. Coupler outputs enter balanced photodiodes (PD), with $I_{out,j} \propto |E_{Laser}||E_{j}^c|$. We note that our scheme relies on maintaining coherency across the entire array, which requires precise optical path lengths and phase shift matchings. In order to adjust for potential phase errors due to process variations or random phase variations, etc., we propose adding a small thermal-phase shifter in each unit cell across the column waveguides.




The photonic elements in our crossbar array present numerous sources of loss, which are critical for system modeling:
\begin{itemize}
    \item Grating coupler: 2 dB~\cite{Luo:18, 45CLO-OFC}
    \item Splitting tree: 0.8 dB~\cite{Zanzi:16}
    \item MMI Crossing Junction - 1.8 dB/junction~\cite{Ma:13}
    \item Waveguide loss: 3 dB/cm~\cite{45CLO-OFC}
    \item Effective loss due to ODAC optical modulation amplitude (OMA) - 4 dB ~\cite{moazeni-jssc2017}
    \item Laser wall-plug efficiency: 15\%
\end{itemize}


\subsection{Peripheral Electronic Circuitry}
While proposed crossbar array can perform optical MAC operation efficiently in a compact footprint with on-chip PCM-based unit cells, the premise of ultra-fast operations relies on energy-efficient and high-speed electro-optical conversions. Here we elaborate on suitable design choices for this aim. We have added area and power estimates of critical peripheral electronics for a 10GHz MAC operation. Note that some assumptions may involve extrapolation from papers reporting work in other processes or operation frequencies. 

\subsubsection{Optical On-chip Transmitter}
As mentioned, the amplitude of input E-field at each row should be modulated with the input vector data. Notice that coherent operation required that phase remain constant with data. This has been previously proposed to be done via MZIs, however, the mm-long footprint of MZIs impose unacceptable area energy overheads for scalable ONN applications. In this design, we propose using micro-ring assisted MZIs (RAMZIs). RAMZI devices has been previously proposed for high linearity~\cite{lipson-RAMZI}, but here we propose using them to perform constant-phase PAM modulation. This can be done by placing one ring-resonator based ODAC in each arm. Such ODACs have been demonstrated in the 45nm silicon photonics with +20GS/s at low power with up to 6-bit accuracy for data modulation~\cite{moazeni-jssc2017}. Each ODAC driver consumes $168fJ$ power and occupies a $0.0012mm^2$ area at 10GS/s~\cite{moazeni-jssc2017}. We also add a thermal tuning over head of $0.72 mW$ per ring-resonator~\cite{moazeni-jssc2017}.
    
\subsubsection{Optical On-chip Coherent Receiver}
Output photocurrents ($I_{out,j}$) should be amplified via a trans-impedance amplifier (TIA) and digitized using an ADC (Fig.~\ref{fig:basic_xbar}). Similar coherent receiver designs have been demonstrated in 45nm silicon photonics with $2.25mW$ per TIA~\cite{mehta-vlsi2019}. ADC power and area at a 10 GHz sample rate is estimated to be $25mW$ and $0.0475mm^2$, respectively, in 45nm CMOS~\cite{zhang202016}.

\subsubsection{SerDes and Clocking}
Operating the ONN crossbar at high clock rates of +10GHz requires a set of serializer/deserializer (SerDes) to interface the optical transmitter and receiver data with the digital backend (SRAM in this case). The serialization ratio depends on the speed of MAC operation and readout speed from SRAM memory. For a 10GHz ONN operation and a $\sim$1GHz backend clock-rate, we assume a 10:1 ratio. Estimated power is roughly $100fJ$ per bit~\cite{moazeni-jssc2017}. The transmit and receive blocks require high speed clock generation and distribution, which we assume will take $200fJ$ power with $0.005mm^2$ per row/column~\cite{moazeni-jssc2017}.
    


\section{Overall Accelerator Architecture}
\label{sec:chip-arch}

\begin{figure}[tb]
    \begin{center}
        \includegraphics[width=0.9\linewidth]{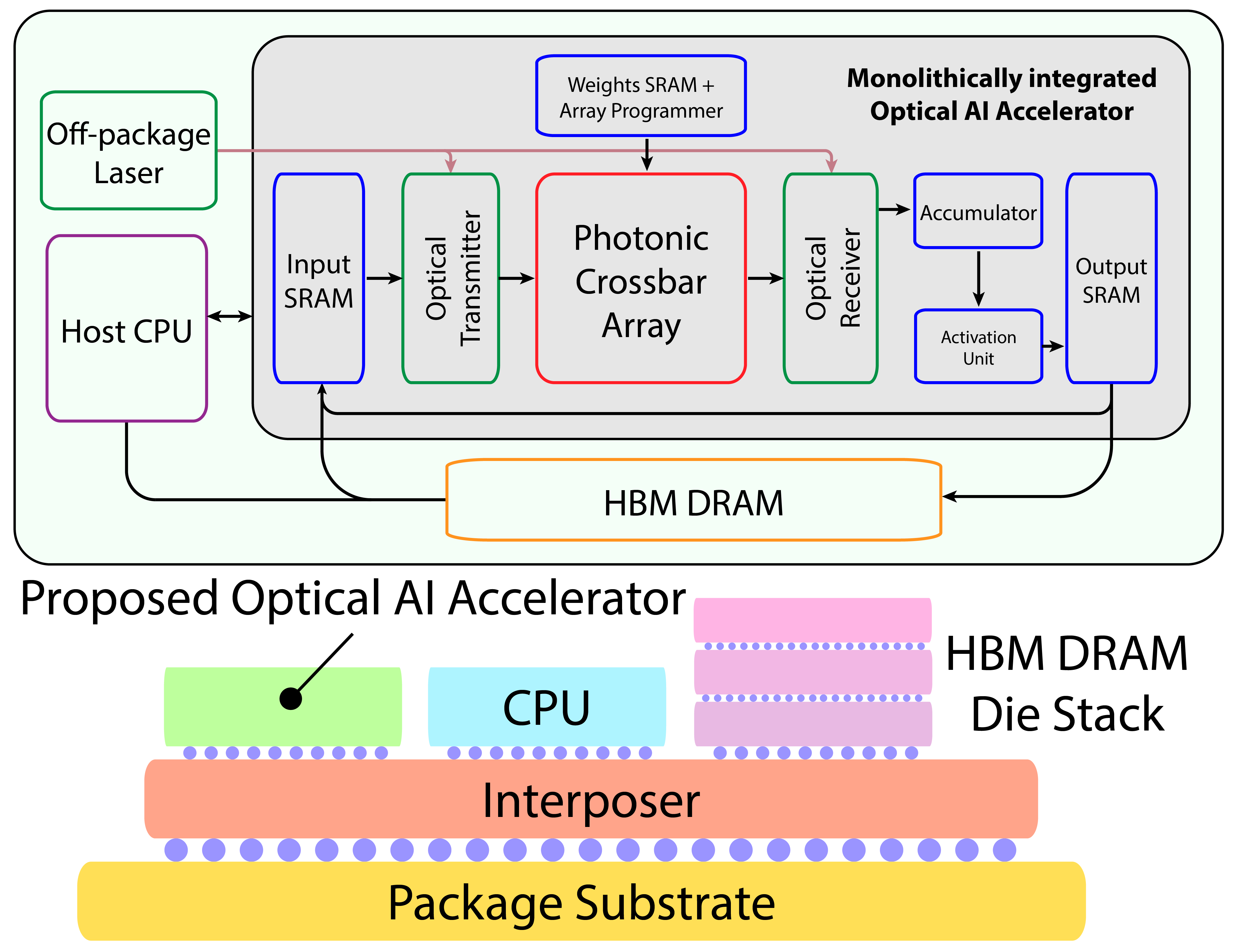}
        \caption{Architecture of proposed photonic crossbar array AI accelerator.}
        \label{fig:chip architecture}
    \end{center}
    \vspace*{-2\baselineskip}
\end{figure}

In this work we focus on CNNs for our modeling and benchmarking, as they are one of the most popular deep learning algorithms due to their strong performance in image classification and object detection~\cite{gu2018recent}. Performing inference using a CNN includes multiple convolutional layers using various sets of filters. Our crossbar array capitalizes on the data reuse inherent to the algorithm by minimizing on- and off-chip data access and executes a CNN as follows: 
The weights in a given 2D filter are flattened and embedded into the crossbar array in the form of PCM programming weights. Since the PCM can only absorb light, all the weights are mapped to a value from between 0 and 1 over 64 levels (6-bits). Once the array has been programmed, the crossbar and peripheral electronics iterate through all subsets of input data (across all batches) and complete proper MAC operations. For any layer, the array will likely have to be programmed multiple times to fully fit all the filters. An accumulator is used at the output of the ADC and deserializer to hold partial sums and add them to new partial sums from the crossbar array. Once complete MAC operations have been computed, they are passed to an activation unit which performs a non-linear transformation to achieve the final output (Fig.~\ref{fig:chip architecture}).


Data can be stored on chip in SRAM blocks (one each for the input, filters, and output) for rapid and low-power access. SRAM area is estimated to be $0.45mm^2$ per $1MB$ in 45nm CMOS~\cite{chen2011power}. On-chip SRAM cannot store the entire datasets and all network parameters for practical CNNs. This necessitates an off-chip DRAM memory, which typically consumes up to $15pJ/b$~\cite{o2017fine} and can dominate the total accelerator power. Here, similar to state-of-the-art AI accelerators, we assume that our chip-scale system can be co-packaged with HBM DRAM stack using advanced packaging~\cite{ayarlabs-intel} (fig.~\ref{fig:chip architecture}), which only consumes $3.9pJ/bit$~\cite{o2017fine}. In order to minimize DRAM access, data can be sent directly from output SRAM to input SRAM at the end of a full layer computation. For modeling, we assumed an SRAM access energy of $50fJ/bit$~\cite{chen2011power}.

Like many non-volatile memories, the PCM in this work has from low programming speed, which can impact crossbar array performance. For instance, PCM programming time can be around $100ns$~\cite{Feldmann2021}, which is $1000\times$ slower than our target 10GHz MAC operation speed. Since the array cannot perform any computations during PCM programming, this has the potential to significantly slow down the effective inference/MAC speed. We propose a "dual core" design featuring two copies of all photonic elements (crossbar array, optical transmitter, and optical receiver) to mitigate this issue. While one core performs computation, the other core can be programmed with the next set of weights. Assuming that programming the array takes less time than running all input data through it, this effectively hides the programming latency, so that computation runs at all times. We note that this solution can be practical only if total area of photonic components will be a small portion of total area. A single laser source can be shared among two crossbar cores. We discuss this scheme further in Section~\ref{sec:results}.

\section{Simulation Framework}
\label{sec:sim-framework}

\begin{figure}[tb!]
    \begin{center}
        \includegraphics[width=0.9\columnwidth]{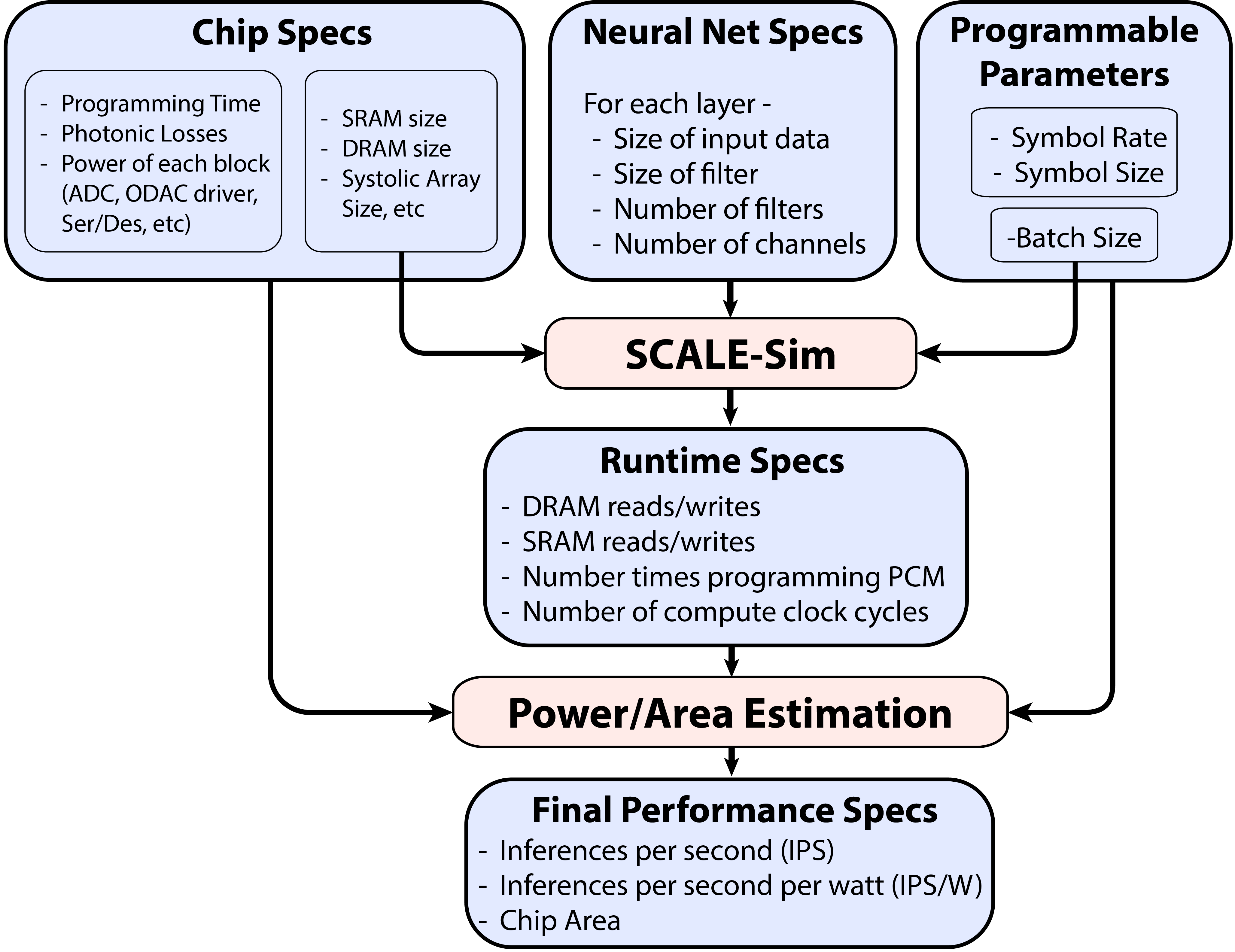}
        \caption{Simulation framework for estimating IPS and IPS/W.}
        \label{fig:simulation framework}
    \end{center}
    \vspace*{-2\baselineskip}
\end{figure}

We modeled our proposed system architecture comprehensively to measure key performance metrics including inferences per second (IPS), inferences per second per watt (IPS/W), and total chip area based on the 45nm monolithic silicon photonics technology. For this aim, we developed a custom simulation framework based on SCALE-sim~\cite{samajdar2018scale} as shown in Fig.~\ref{fig:simulation framework}. There are two main steps in estimating the key performance metrics values:

\textbf{(1) Calculating the runtime specs:} This includes the number of SRAM/DRAM accesses, crossbar programming cycles, and number of MAC compute cycles.

\textbf{(2) Calculating high-level metrics:} This is estimated using the runtime specs as well as the specs of all chip components - programming time, photonic losses, ADC power, DRAM access power, etc. (See Sections~\ref{sec:xbar-mvm} and ~\ref{sec:chip-arch})

Calculating runtime specs must be done for a specific neural network and dataset. We achieve this using SCALE-Sim, a cycle-accurate tool for simulating CNN accelerators~\cite{samajdar2018scale}. We made slight modifications to SCALE-Sim to account for non-unity batch sizes, the presence of an accumulator, and output SRAM data reuse.

In step two of our framework, the number of array programming cycles and MAC  compute operation cycles are combined with MAC operation speed and programming time to calculate inference time and corresponding IPS. Total chip power is calculated based on clock speed, number of SRAM and DRAM accesses (from step one), plus the energy of optical and electrical components. Chip area is estimated similarly from SRAM sizes, digital accumulator and activation block areas, photonic devices footprints, and area of peripheral electronics, including ADCs, DACs, and clocking.


\section{System-Level Performance Trends and Optimization}
\label{sec:results}

Our holistic simulation framework enables us to search the many-dimensional chip design space to find the optimal configuration for ResNet50 inference. We do so by studying the trends in terms of IPS and IPS/W for various design parameters. Here we present observed critical trends and our approach to choosing the optimal design. While we investigate the effect of array size, batch size, dual core scheme, and SRAM size, we hold a constant MAC operation speed of $10GHz$, since the power of many of the peripheral electronics (specifically ADCs) may rise steeply at higher speeds.

\begin{figure}[tb!]
    \begin{center}
        \includegraphics[width=0.65\linewidth]{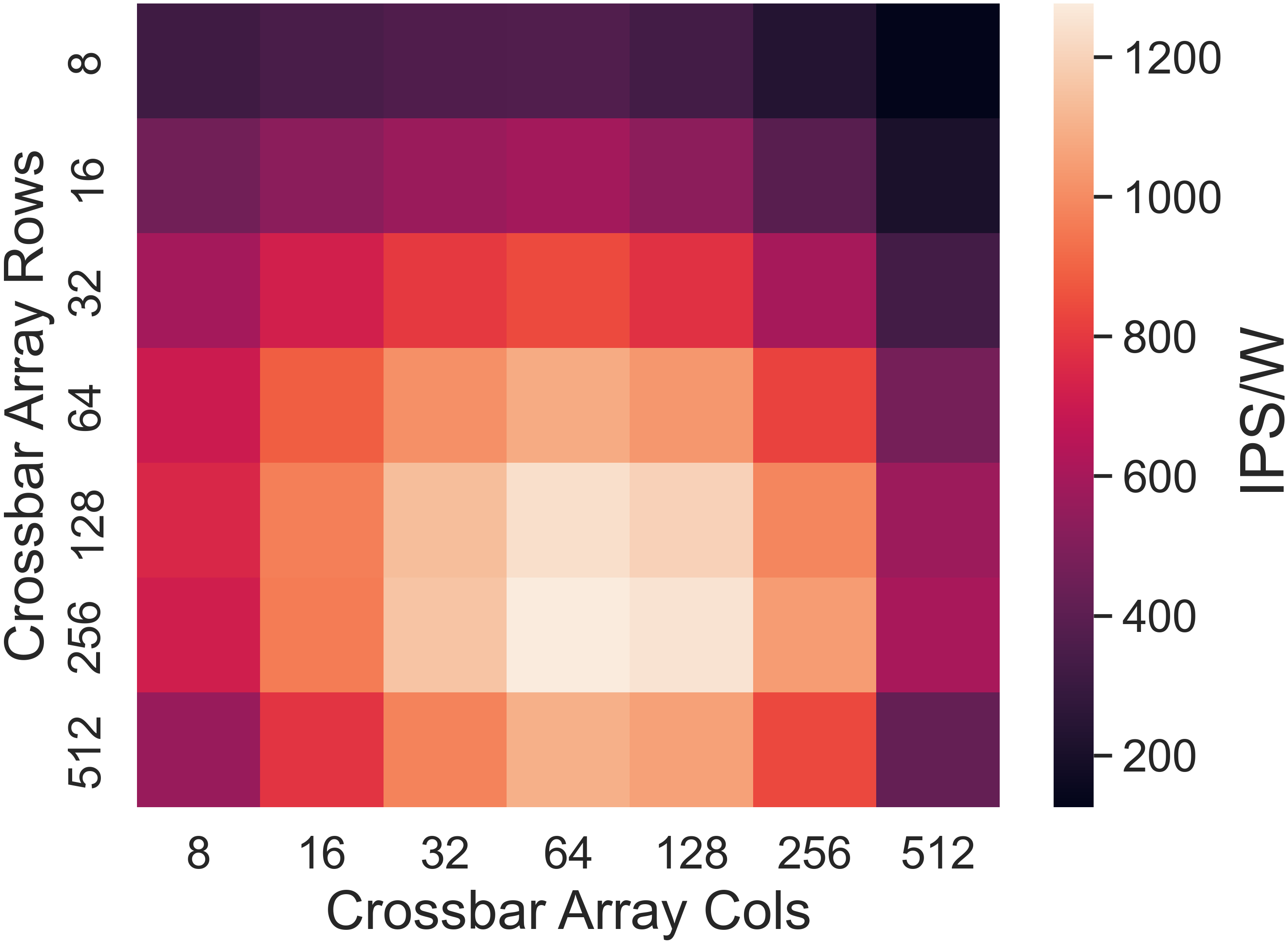}
        \caption{IPS/W as a function of crossbar rows and columns.}
        \label{fig:IPSW rows cols}
    \end{center}
    \vspace*{-2\baselineskip}
\end{figure}

\subsection{Trends in Array Performance}
Major trends are described below. All figures and data referenced in this section assume the following default chip parameters except when clearly being swept: Array size: 32 x 32, SRAM sizing: 26.3 MB (input), 0.75 MB (output), 0.75 MB (filter), and 0.75 MB (accumulator), dual-core architecture, and batch size: 32.

\subsubsection{Dual vs. Single Core} Dual core design increases the IPS, but power consumption is also consistently higher since computing and programming happen simultaneously. As a result, IPS/W is the same regardless of the core count. 

\subsubsection{Array Size} IPS always increases approximately linearly with the array size ($N \times M$). This is no surprise because a larger array can process larger matrices/vectors, so less programming and compute cycles are required to reach the final sum. However, the effect of array size on IPS/W is more complicated. As the array size increases, the power of peripheral electronics (such as ODAC drivers or ADCs) grows. Also, higher laser power is required for larger arrays because the power input to each row must spread out to more columns and rows. However, both of these effects increase the power less than linearly since many components are unaffected (such as DRAM accesses).
However, photonic losses in the crossbar array, such as those due to MMI waveguide crossings or waveguides themselves, grow exponentially. Ultimately this causes power to grow more than linearly with array size, so IPS/W drops. We observe a peak IPS/W at array size of 128-256 rows and 64-128 columns (Fig.~\ref{fig:IPSW rows cols}). 

\subsubsection{Batch Size and SRAM Size} 
\begin{figure}[tb!]
    \begin{center}
        \includegraphics[width=\linewidth]{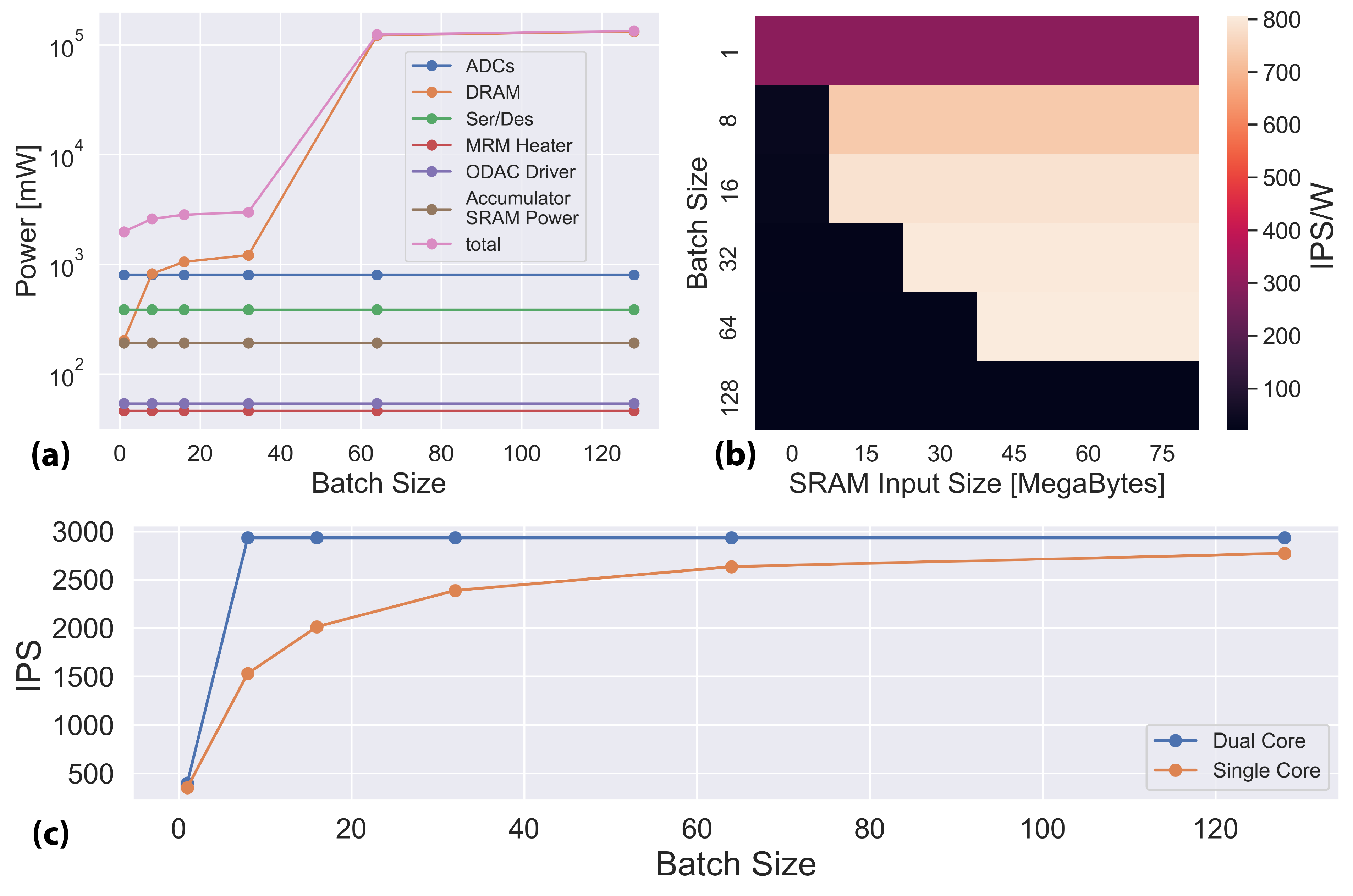}
        \caption{(a, b) Effect of the SRAM size and batch size on chip power and IPS/W, (c) Effect of dual-core architecture on relationship between batch size and IPS.}
        \label{fig:batch SRAM}
    \end{center}
    \vspace*{-2\baselineskip}
\end{figure}

Running CNNs with large batch sizes offers the possibility of amortizing the cost of array programming time over a larger set of input data, increasing IPS. However, if not all of the input data across all batches can fit on the input SRAM, data will routinely have to be loaded on and off the chip every time the array is reprogrammed. This effect can be seen in~\ref{fig:batch SRAM}a, where DRAM access energy rises steeply between batch sizes of 32 and 64. For any batch size there is a critical input SRAM size that must be met for optimal IPS/W, and any larger input SRAM size does not increase the performance further (Fig.~\ref{fig:batch SRAM}b). 

It is important to note that the area overhead of increasing SRAM size is significant due to the chip area limitations. While smaller batch size is more favorable from this perspective becuase it reduces the need for large SRAM, it can limit IPS. This can be alleviated using our dual-core scheme which can hide the array's programming latency, so IPS becomes only a function of computation time. Fig.~\ref{fig:batch SRAM}c shows the large increase in IPS a dual-core architecture can have for small-batch systems. 

\subsection{Proposed Approach for Optimization} Building on these observed trends, we develop the following flow for optimizing system performance: Firstly, we find the smallest batch size that is large enough to avoid a low IPS caused by a large programming time that cannot be hidden by our dual core architecture. We do not need to be overly concerned with increasing the batch size beyond this minimal amount because it will have minimal effect on IPS/W, even for infinitely large SRAM (Fig.~\ref{fig:batch SRAM}b). Next, we maximize the SRAM size without exceeding a practical chip size ($\sim1cm^2$ in here), assuming that SRAM dominates chip area, validated in Fig.~\ref{fig:power area}. Finally, we find the the array size that maximizes IPS/W. When there are multiple array sizes with similar IPS/W, we pick that largest array as it yields a higher IPS, even though exact optimal size may depend on the CNN.

\section{Results and Comparison to State-of-the-Art}
\label{sec:compare-soa}

Using our optimization framework, we arrive at the following system parameters for optimal performance of our ONN, with dual-core architecture and a $10 GHz$ MAC operation speed.

\begin{itemize}
    \item Array size: 128 rows $\times$ 128 columns
    \item SRAM sizes: 26.3 MB (input), 0.75 MB (output), 0.75 MB (filter), and 0.75 MB (accumulator)
    \item Batch Size: 32
\end{itemize}

\begin{figure}
    \begin{center}
         \includegraphics[width=0.95\linewidth]{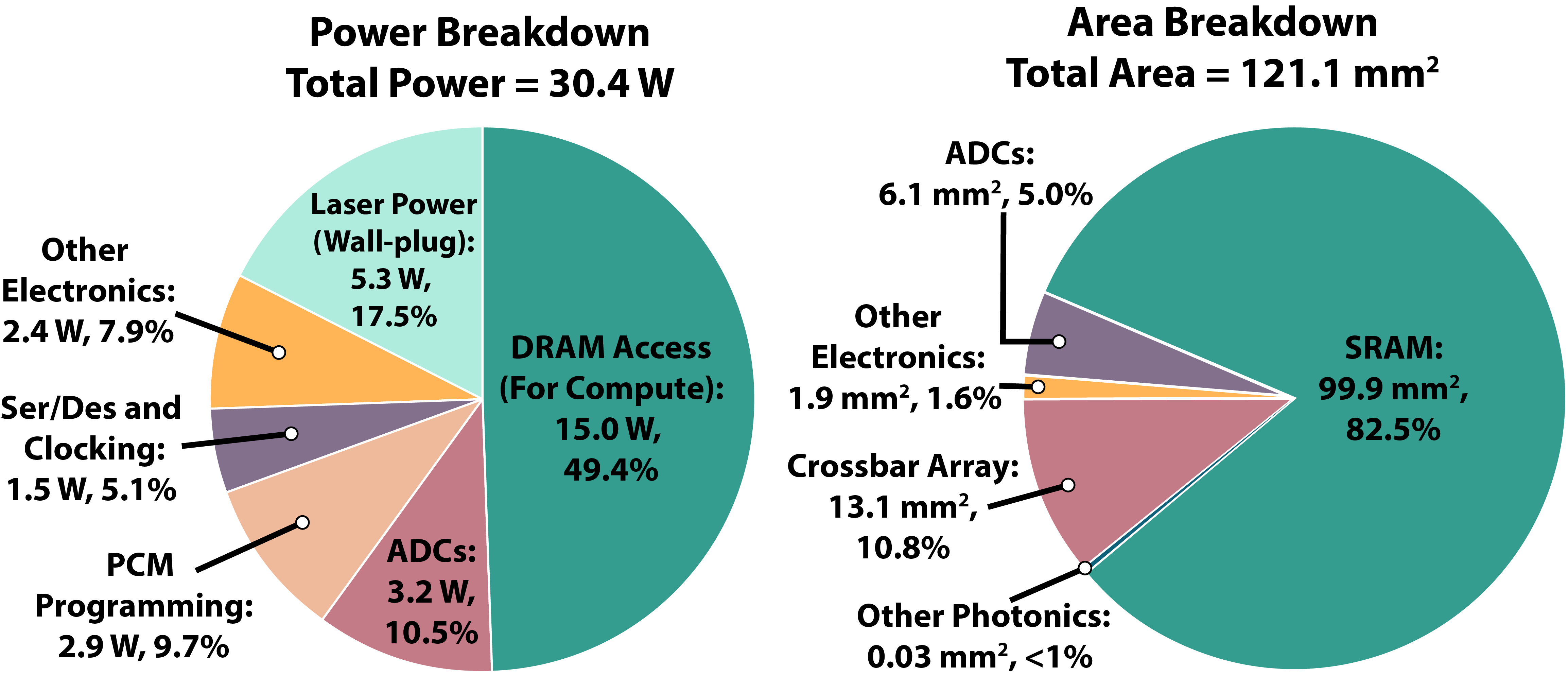}
        \caption{Power and area breakdown of our proposed accelerator.}
        \label{fig:power area}
    \end{center}
    \vspace*{-2\baselineskip}
\end{figure}

The performance of this system is compared with the state-of-the-art NVIDIA A100 GPU (INT8 mode with a batch of 128) for ResNet50 in the table below~\cite{nvidia_developer_2022}. Our proposed system achieves similar IPS with $15.4\times$ lower power and $7.24\times$ lower area. Overall power and area breakdowns are illustrated in Fig.~\ref{fig:power area}. While the power consumption is dominated by DRAM accesses, the area is mainly dominated by the SRAM blocks.

\renewcommand*{\arraystretch}{2}
\begin{table}[h]
\begin{center}
    \footnotesize
    \begin{tabular}{ | l |l | l |l | l|}
    \hline
    \textbf{System} & \textbf{IPS} & \textbf{IPS/W}   & \textbf{Power} & \textbf{Area} \\ \hline
    This work & 36382 &  1196 & 30W & 121 $mm^2$ \\ \hline
    Nvidia A100 & 29,733 &	75  & 396W & 826 $mm^2$ \\ \hline
    \end{tabular}
    \medskip
\end{center}
\vspace*{-2\baselineskip}
\end{table}

\section{Conclusion}
\label{sec:conclusion}

We have presented an optical AI accelerator using the coherent crossbar design with on-chip PCM weight storage. The key performance metrics of this design are holistically modeled and estimated using practical and realistic reported results based on 45nm monolithic silicon photonic technology. We have discussed our modeling and system optimization framework. Our approach can be used for optimizing other types of emerging AI accelerators as well. Furthermore, we proposed a dual core system design that can hide programming latency. Our results show that overall system performances greater than the state-of-the-art can be achieved. This work can be a first step towards designing and optimizing large-scale ONN accelerators for real-world applications by considering all the crucial aspects including SRAM/DRAM memory, electronic-photonic integration, packaging, and all peripheral electronics.

\bibliographystyle{IEEEtran}
\bibliography{emit-refs}

\end{document}